\documentclass[floatfix,twocolumn,prb, aps,groupedaddress,10pt,longbibliography]{revtex4-1}
\usepackage[english]{babel}
\usepackage[colorlinks=true,linkcolor=blue, citecolor=blue, urlcolor= blue]{hyperref} 
\usepackage{amsmath}
\usepackage[caption=false]{subfig}
\usepackage{dcolumn}
\usepackage{color}
\usepackage{graphicx} 
\usepackage{standalone}
\usepackage{tikz} 

\definecolor{rubblue}{HTML}{003E64}

\newcommand{\sigmaint}{\mat{\sigma}^{\text{\scriptsize int}}}
\newcommand{\sigmabulk}{\mat{\sigma}^{\text{\scriptsize bulk}}}
\newcommand{\musphere}{\ensuremath{\mu_}{\text{\scriptsize I}}}
\newcommand{\muouter}{\mu_{\text{\scriptsize M}}}
\newcommand{\lasphere}{\lambda_{\text{\scriptsize I}}}
\newcommand{\laouter}{\lambda_{\text{\scriptsize M}}}

\newcommand{\Pa}{\text{Pa}}
\newcommand{\force}{\vec{f}}
\newcommand{\forcebulk}{\vec{f}^{\text{\scriptsize bulk}}}
\newcommand{\forcepressure}{\vec{f}^{\text{\scriptsize bulk-p}}}
\newcommand{\forcedeviatoric}{\vec{f}^{\text{\scriptsize bulk-dev}}}
\newcommand{\forceint}{\vec{f}^{\text{\scriptsize int}}}
\newcommand{\intenergy}{\sigma}
\newcommand{\intenergyab}{\sigma_{\alpha\beta}}

\renewcommand{\pl}{p_{\text{l}}}
\newcommand{\pv}{p_{\text{v}}}
\newcommand{\Vl}{V_{\text{l}}}
\newcommand{\Vv}{V_{\text{v}}}
\newcommand{\mat}[1]{\boldsymbol{#1}}
\newcommand{\diff}{\text{d}}
\renewcommand{\vec}[1]{\boldsymbol{\mathbf{#1}}}

\usepackage{etoolbox} 
\usepackage{breqn} 
\usepackage{flexisym} 
\usepackage{lineno}

\BeforeBeginEnvironment{dmath}{\begin{nolinenumbers}}
\AfterEndEnvironment{dmath}{\end{nolinenumbers}}
\BeforeBeginEnvironment{dmath*}{\begin{nolinenumbers}}
\AfterEndEnvironment{dmath*}{\end{nolinenumbers}}

\BeforeBeginEnvironment{figure}{\begin{nolinenumbers}}
\AfterEndEnvironment{figure}{\end{nolinenumbers}}
\BeforeBeginEnvironment{figure*}{\begin{nolinenumbers}}
\AfterEndEnvironment{figure*}{\end{nolinenumbers}}

\begin{document}

\title{A multi-phase-field method for surface tension-induced elasticity}

\author{Raphael Schiedung}
\email{raphael.schiedung@rub.de}

\author{Ingo Steinbach}

\author{Fathollah Varnik}
\email{fathollah.varnik@rub.de}

\affiliation{Ruhr Universit\"at Bochum$,$ Interdisciplinary Center for Advanced
Materials Simulation (ICAMS)$,$ Universit\"atsstr.~150$,$ 44801 Bochum$,$
Germany}

\date{\today}

\begin{abstract}
    A consistent treatment of the coupling of surface energy and elasticity within the multi-phase-field framework is presented.
    The model accurately reproduces stress distribution in a number of analytically tractable, yet non-trivial, cases including different types of spherical heterogeneities and a thin plate suspending in a gas environment.
    It is then used to study the stress distribution inside elastic bodies with non-spherical geometries, such as a solid ellipsoid and a sintered structure.
    In these latter cases, it is shown that the interplay between deformation and spatially variable surface curvature leads to heterogeneous stress distribution across the specimen.
\end{abstract}


\maketitle

\section{Introduction}

The issue of surface or interface energy-induced stress has long been recognized as a fundamental research topic, and has been investigated already by J.~Willard Gibbs in the early years of the past century~\cite{gibbs_scientific_1906}.
Later, a differentiation was introduced between the surface energy, the work necessary to create a unit area of surface, and the surface tension, the tangential stress (force per unit length) in the surface layer.
This distinction has led to the so-called Shuttleworth equation which relates the surface tension stress to the surface energy~\cite{shuttleworth_surface_1950}.

In the decades following this seminal work, several authors addressed the issue of elasticity and the surface or interface tension of solid surfaces and interfaces~\cite{gurtin_continuum_1975, gurtin_addenda_1975, cammarata_surface_1994, cammarata_surface_1994-1, gurtin_general_1998,kramer_note_2007}.
Alternative interpretations of the Shuttleworth's equation have been also proposed with regard to a rigorous distinction of and mutual relation between the surface stress, surface tension and surface energy~\cite{bottomley_alternative_2001, makkonen_misinterpretation_2012}.

On the application side, the advent of piezomagnetic, ferroelastic and piezoelectric nanomaterials~\cite{yan_modified_2017} renewed the interest in a better understanding of surface energy-induced stresses and their role for the materials' functionality.
Indeed, at these small scales, the action of surface tension can lead to considerable internal stresses and elastic deformation of the material with important size effects~\cite{sharma_effect_2003,sharma_size-dependent_2006}.
Surface or interface energy-induced stresses are also of fundamental interest for all elasticity-related interface phenomena such as phase transformation kinetics~\cite{fischer_role_2008, slutsker_phase-field_2008,levitas_size_2011, levitas_phase_2016}, surface or interface diffusion, and thin film-related phenomena~\cite{lu_dynamics_2001, lu_symmetry_2002, lu_patterning_2004}.

In this context, a computational approach which allows to account, under complex geometries, for the effect of surface and interface energy on the stress and deformation state at the nanoscale is highly desirable.
Motivated by this idea, we build upon the well-established multi-phase-field method~\cite{steinbach_phase-field_2009, steinbach_phase-field_2013} and present a consistent formulation for the coupling between surface energy and elasticity in solid bodies of arbitrary shape and therewith provide an efficient new numerical model for the study of complex mechanical deformation and heterogeneous stress states in nanomaterials.

The proposed methodology has important new applications on the nanoscale, where the heterogeneous deformation and stress state of the material largely determines its electric or magnetic response, as is the case, e.g.~in ferroelastic, piezoelectric and piezomagnetic phenomena.

After a brief introduction in Sec.~\ref{sec:surface_energy}, the proposed coupling of elasticity and surface energy into the multi-phase-field framework is presented in Sec.~\ref{sec:model} of this paper.
In Sec.~\ref{sec:results}, the method is validated using a number of non-trivial test cases for which analytic solution can be obtained.
The thus established method is then used to investigate the behavior of stress within bodies of complex shape such as an ellipsoid and a sintered structure.
The existence of heterogeneous stress states in these structures is studied and its connection to local curvature is discussed.
A summary of the main results is provided in Sec.~\ref{sec:conclusion}.

\section{Surface energy and elasticity}
\label{sec:surface_energy}

Let us first consider a liquid phase coexisting with its vapor.
For a quasi-static process at equilibrium, the total differential of the free energy reads
\begin{dmath}
     \diff F
    = \intenergy \, \diff A
    - \pl \, \diff \Vl
    - \pv \, \diff \Vv
    \label{eq:dF-liq}
\end{dmath},
where $\pl$ and $\pv$ denote pressure in the liquid and vapor phases, respectively.
$\Vl$ and $\Vv$ are the volumes of respective domains.
$A$ is the surface area and $\intenergy$ the surface energy.
We assume that the condition of local thermodynamic equilibrium is satisfied so that the free energy $F$ is always at its minimum value.
This implies that a variation of control parameters such as volume of the liquid domain will leave $F$ unchanged to the first order, $ \diff F / \diff A =0$.
Using $ \diff \Vv=- \diff \Vl$, one then obtains from Eq.~(\ref{eq:dF-liq})
\begin{dmath}
    \pl - \pv = \intenergy \frac{ \diff A}{ \diff \Vl}
\label{eq:dp-LV}
\end{dmath}.

For a spherical drop of radius $r$ in $d$ dimensions, $ \diff A/ \diff \Vl=(d-1)/r=\kappa$, where $\kappa$ denotes the surface curvature.
Thus, one recovers the Young-Laplace equation for the pressure difference $\Delta p \equiv \pl - \pv$,
\begin{dmath} \label{eq:young_laplace}
    \Delta p =\dfrac{(d-1)\sigma}{R} = \intenergy \kappa
\end{dmath}.

In the case of solid surfaces or interfaces, the material may deform elastically giving rise to internal stresses.
In this case, the hydrostatic pressure is no longer necessarily constant inside the bulk of the material but depends both on surface tension-induced forces and the internal deformation state.

In the special case of two elastic phases $\alpha$ and $\beta$ at contact, the total differential of free energy reads
\begin{dmath} \label{EqdFSolid}
     \diff F
    = \intenergyab \, \diff A
    + V_{\alpha} \mat{\sigma}_{\alpha} \operatorname{:} \diff \mat{\varepsilon}_{\alpha}
    + V_{\beta} \mat{\sigma}_{\beta} \operatorname{:} \diff \mat{\varepsilon}_{\beta}
\end{dmath},
where $\intenergyab$ is the surface or interface energy per unit area and $\operatorname{:}$ denotes the double inner product ($\mat A \operatorname{:} \mat B = \sum_{i,j}A_{ij}B_{ji}$).
$\mat{\sigma}_{\alpha}$ and $\mat{\sigma}_{\beta}$ are the stress tensors with the conjugate strains $\mat{\varepsilon}_{\alpha}$ and $\mat{\varepsilon}_{\beta}$ of the phases $\alpha$ and $\beta$, respectively.
$V_\alpha$ and $V_\beta$ are the undeformed volumes of the corresponding phases.

It can be shown that $\intenergyab$ couples to the elastic deformation as well.
For this purpose, we consider $\intenergyab$ as energy per unit area in the deformed configuration $A$.
An undeformed surface or interface of the area $A_0$ would be deformed into the area $A$ by a strain alongside the interface $\mat \varepsilon_{\mathrm{I}}$ with
$A = A_0 \left(1 + \mat 1 \operatorname{:} \mat {\varepsilon}_{\mathrm{I}}\right)$.
The total differential of free energy is now written as,
\begin{dmath}\label{EqdFde}
     \diff F
    =\intenergyab \diff A_0 +
    A_0 \intenergyab \mat 1 \operatorname{:} \diff \mat{\varepsilon}_{\mathrm{I}}
    + V_{\alpha} \mat{\sigma}_{\alpha} \operatorname{:} \diff \mat{\varepsilon}_{\alpha}
    + V_{\beta} \mat{\sigma}_{\beta}\operatorname{:} \diff \mat{\varepsilon}_{\beta}
\end{dmath}.
Equation~(\ref{EqdFde}) tells us that, starting from a stress free state ($\mat\sigma_\alpha=\mat\sigma_\beta=\mat 0$) the system may deform along the interface to allow a reduction of surface area.
Interestingly, as will be shown below (see section~\ref{sec:plate}), this may occur also in the absence of curvature, i.e.~for a planar interface, leading to a non-isotropic stress tensor.

In general, the surface or interface energy of solids $\intenergyab$ is a function of temperature, crystal lattice orientation and elastic deformation of the surface or interface~\cite{gurtin_continuum_1975, gurtin_addenda_1975, cammarata_surface_1994, cammarata_surface_1994-1, gurtin_general_1998,kramer_note_2007}.
In the case of solid-solid interfaces, the dependency of the interface energy on the elastic deformation might even be different for both sides of the interface~\cite{cammarata_surface_1994, cammarata_surface_1994-1}.
For the sake of simplicity, however, this work focuses on the case of constant surface or interface energy.

Equation~(\ref{EqdFde}) can easily be extended to account for the presence of $N$ thermodynamic phases and their mutual interfaces,
\begin{eqnarray}\label{EqdFdeMulti}
     \diff F &= &\sum_{\alpha, \beta > \alpha}^N \left (\intenergyab \diff A_0
    + A_{0,\alpha\beta} \intenergyab \mat 1 \operatorname{:} \diff \mat{\varepsilon}_{\alpha\beta}\right )\nonumber \\
    & &+ \sum_{\alpha=1}^N V_{\alpha} \mat{\sigma}_{\alpha} \operatorname{:} \diff \mat{\varepsilon}_{\alpha}.
\end{eqnarray}
In Eq.~(\ref{EqdFdeMulti}), the multi-phase version of interface energy $\intenergyab$ and the deformation tensor $\mat \varepsilon_{\alpha\beta}$ corresponding to the interface between the phases $\alpha$ and $\beta$ are used.
For the sake of brevity, we also use
$\sum^N_{\alpha, \beta > \alpha} \equiv \sum_{\alpha=1}^N\sum_{\beta=\alpha+1}^N
$.

The above description is based on the so-called sharp interface picture, where bulk phases are separated by infinitely thin interfaces.
Within the multi-phase-field method, which builds the mathematical framework for the present model, sharp interfaces are replaced by diffuse ones, assigning a finite thickness to the interface domain.
One of the main advantages of this description is that interface tracking is no longer necessary.
Moreover, problems aroused by discontinuous changes of physical quantities at sharp interfaces are avoided.

\section{The phase-field model}
\label{sec:model}
The main motivation for the development of the present method is the need to account for the effect of surface or interface energy on the deformation state and the corresponding stress distribution in nano-structured materials.
In order to describe the associated complex geometries of coexisting gas and multiple solid phases, the so-called multi-phase-field method provides a good approach.
The next section introduces the proposed method with a particular focus on the new aspect, the coupling between the surface or interface energy and the mechanical deformation in the elastic regime.

\subsection[Interface energy functional]{The density functional for free energy}
A detailed account of the multi-phase-field method can be found in~\cite{steinbach_phase-field_2009, steinbach_phase-field_2013} and references therein.
For the sake of completeness, however, a brief introduction to this fast evolving field is given below.

A basic ingredient of any phase-field method is the so-called phase-field function $\phi_\alpha(\vec{r})$.
This function can be viewed as the fraction of the volume element $ \diff \vec r^3$ around $\vec r$, occupied by the thermodynamic phase $\alpha$.
As a consequence, $\sum_{\alpha} \phi_{\alpha}(\vec r)=1$.
Once all phase-field functions are known at a given point in space, physical quantities at that point can be evaluated as average over thermodynamic phases present at that point.
For example, the density is given by $\rho(\vec r)=\sum_{\alpha} \rho_\alpha \phi_\alpha(\vec r)$.
Given the set of phase-fields, the physical model of interest is constructed starting from a free energy functional $F$,
\begin{widetext}
\begin{dmath} \label{eq:energy_functional}
    F\left[\left\{\phi_{\alpha} \right\}, \vec{u}\right]
    =
    \int_{\Omega} \diff^3\vec r \,
    f \left(\left\{\phi_{\alpha}(\vec r)\right\}, \left\{\nabla \phi_{\alpha}(\vec r) \right\}, \vec u(\vec r), \nabla \vec u(\vec r) \right).
\end{dmath}
\end{widetext}
We consider $F$ to be a functional of $N$ phase-fields, $\phi_\alpha$ with $\alpha \in \left[1,N\right]$, and the displacement vector field $\vec u$.
The free energy density, $f$, in Eq.~(\ref{eq:energy_functional}) is usually split into individual parts,
\begin{dmath} \label{eq:energy_density}
    f
    =
    \sum_{\alpha,\beta > \alpha}^N
    f^{\text{interface}}_{\alpha \beta} + f^{\text {elastic}}
\end{dmath},
where $f^{\text{interface}}_{\alpha \beta}$ is the energy density of the interface between phases $\alpha$ and $\beta$ and $f^{\text{elastic}}$ is the elastic energy density associated with the deformation state of the material.
In usual phase-field models of phase transformation kinetics, there is also another important term which accounts for the contribution of the thermodynamic bulk phase, $\sum_{\alpha=1}^{N}f^{\text{bulk}}_{\alpha}$.
This term is important in the presence of thermodynamic driving forces for phase transformation, such as temperature or solutal undercooling~\cite{steinbach_phase-field_2013}.
However, since the focus of the present study is on the deformation induced by surface or interface energy, it is neglected here.
In order to proceed further, we need closed expressions for the first and second terms on the right hand side of Eq.~(\ref{eq:energy_density}).
The elastic energy density is given by
\begin{dmath} \label{eqn:sum_engery_density_elas}
    f^{\text{elastic}} =
    \frac{1}{2} \mat \varepsilon \operatorname{:} \mat{C} \operatorname{:} \mat \varepsilon,
\end{dmath}
with the strain tensor $\mat \varepsilon$ and the stiffness tensor $\mat C$.
The deformation tensor is obtained from the spatial gradient of the displacement field via
$\mat \varepsilon = \frac{1}{2} \left( \nabla \vec u + (\nabla \vec u)^T \right)$, where the superscript $T$ stands for the transpose operator.
The stiffness tensor is not constant but depends on the distribution of phases in space.
For simplicity, we use a weighted average of stiffness constants of the available phases,
$\mat{C} \left(\vec x \right) = \sum_\alpha^N \phi_\alpha \left( \vec x \right) \mat{C}_\alpha$.

For the interface free energy density, a diffuse interface analog of $\mat \varepsilon_{\alpha\beta}$ (see Eq.~(\ref{EqdFdeMulti})) is introduced via the projection of the strain field onto the interface.
This is achieved by using the projection matrix $\mat P_{\alpha\beta}=\mat 1 - \vec{n}_{\alpha\beta} \vec{n}_{\alpha\beta}$, where $\vec{n}_{\alpha\beta}$ is the unit normal vector between phases $\alpha$ and $\beta$, given by
\begin{dmath} \label{eq:interface_normal}
    \vec{n}_{\alpha\beta}
    =
    \frac{\phi_\alpha \nabla \phi_\beta - \phi_\beta \nabla \phi_\alpha}
    {\left\Vert \phi_\alpha \nabla \phi_\beta
        - \phi_\beta \nabla \phi_\alpha \right\Vert}
\end{dmath}.

The interface free energy density associated with phases $\alpha$ and $\beta$ thus reads,
\begin{dmath} \label{eq:energy_density_intf}
    f^{\text{interface}}_{\alpha \beta}
    =
    I_{\alpha \beta}
    \intenergy_{\alpha \beta}
    \left( \mat 1 + \mat P_{\alpha \beta} \operatorname{:} \mat{\varepsilon} \right)
\end{dmath},
where $I_{\alpha \beta}$ is a characteristic or indicator function for the interface between the phases $\alpha$ and $\beta$,
\begin{dmath} \label{eq:idicator}
    I_{\alpha \beta}
    =
    \frac{4}{\eta}
    \left[
        - \frac{\eta^2}{\pi^2}
        \nabla \phi_{\alpha} \cdot \nabla \phi_{\beta}
        +
        \left\vert \phi_{\alpha} \right\vert \left\vert \phi_{\beta} \right\vert
    \right]
\end{dmath}.
As the name already suggests, $I_{\alpha \beta}$ is only non-zero in the interface between phases $\alpha$ and $\beta$.

The interfacial free energy density $ f^{\text{interface}}_{\alpha \beta}$ is commonly used in multi-phase-field methods~\cite{steinbach_multi_2006, steinbach_phase-field_2009, steinbach_phase-field_2013}.
The major novelty of the present work is the introduction of the interface-deformation term $\mat P_{\alpha \beta}\operatorname{:}\mat \varepsilon$ in Eq.~(\ref{eq:energy_density_intf}).
This term accounts for the free energy change arising from an elastic deformation alongside the surface or interface.

\subsection{Dynamic equations and mechanical equilibrium}
The time evolution of the phase-field function $\phi_\alpha$ is obtained from the dissipative ansatz~\cite{steinbach_phase-field_2009}
\begin{dmath*}
\frac{\partial \phi_\alpha}{\partial t}
=- M_\phi \sum_{\beta \neq \alpha} \Big[\frac{\delta F}{\delta \phi_\alpha} - \frac{\delta F}{\delta \phi_\beta}\Big],
\end{dmath*}
where $M_\phi$ is a mobility coefficient.
In the elastic limit considered in this work, the dynamics of displacement field $\vec u$ is governed by momentum conservation and is of the second order with respect to time derivative,
\begin{dgroup}
\begin{dmath} \label{eqn:u_dynamics}
    \frac{d^2 \vec{u}}{ d t^2} = M_{u} \nabla \cdot \mat\sigma\,\,\,\ \textrm{with}
\end{dmath}
\begin{dmath} \label{eq:stress-tensor-def-dF}
    \nabla \cdot \mat\sigma = \frac{\delta F}{\delta \vec{u}}
\end{dmath}.
\end{dgroup}
$M_u$ is a kinetic coefficient of the dimension of inverse mass and $\mat{\sigma}$ is the stress tensor.
Equation~(\ref{eqn:u_dynamics}) is equivalent to the Cauchy momentum equation, reformulated for the displacement instead of the velocity field.

In many cases of interest, the dynamics of the phase-field function, described by $\dot\phi_\alpha$, is slow compared to the speed of elastic deformation.
Therefore, on the relatively long time scale associated with the evolution of $\phi_\alpha$, the displacement field can reach local mechanical equilibrium between the two subsequent updates of the phase-field function.
This means that, the equation for $\vec u$ can be solved assuming no acceleration, $ d^2 \vec u / d t^2 = 0$.
Within this approximation, Eq.~(\ref{eqn:u_dynamics}) simplifies to $\nabla \cdot\mat{\sigma} = \vec 0$.
In the remaining of this paper, we will focus on this interesting limit and will work out its consequences for the mechanical behavior of solid bodies on the nanoscale, where interface effects play a major role.

Recalling the definition of the free energy functional $F$, and the elastic and interface free energy densities, Eqs.(\ref{eqn:sum_engery_density_elas}) and (\ref{eq:energy_density_intf}), one obtains from Eq.~(\ref{eq:stress-tensor-def-dF}) at mechanical equilibrium,
\begin{dmath} \label{eq:mech-equil}
    \nabla \cdot
    \left[
        \mat{C}
        \operatorname{:} \mat \varepsilon
      + \sum_{\alpha, \beta > \alpha}^N I_{\alpha \beta} \intenergy_{\alpha \beta}\mat P_{\alpha\beta}
    \right]=\vec 0
\end{dmath}.
It is seen from Eq.~(\ref{eq:mech-equil}) that the total stress tensor has two main contributions,
\begin{dmath} \label{eq:sigmatotal}
\mat \sigma = \sigmabulk + \sigmaint,
\end{dmath}
with
\begin{dgroup}
\begin{dmath} \label{eq:sigmabulk}
    \sigmabulk = \mat C \operatorname{:} \mat \varepsilon \;\;\;\mathrm{and}
\end{dmath}
\begin{dmath} \label{eq:sigmastar}
    \sigmaint = \sum_{\alpha, \beta > \alpha}^N I_{\alpha \beta} \intenergy_{\alpha \beta} \mat P_{\alpha\beta}
\end{dmath}.
\end{dgroup}

Please note that $\sigmaint$ has the dimension of energy per unit volume and is thus similar to the ordinary stress tensor.
This is a consequence of the finite interface thickness so that a volume rather than a surface area is associated with the interface domain.
Nevertheless, as seen from the presence of the indicator function, $I_{\alpha\beta}$ in Eq.~(\ref{eq:sigmastar}), the action of $\sigmaint$ is restricted to the interface domain.

The projection operator $\mat P_{\alpha\beta}$ on the other hand, ensures that $\sigmaint$ has only tangential components along the interface.
The magnitude of $\sigmaint$ is controlled by the surface or interface energy $\intenergy_{\alpha\beta}$.
Interestingly, the interface stress tensor $\sigmaint$ leads to a force along the normal direction that is proportional to the curvature.
This is easily seen by applying the divergence operator to Eq.~(\ref{eq:sigmastar}).
Using the definition of the projection operator in terms of the interface normal vector $\vec n_{\alpha\beta}$, this yields
\begin{dmath}
    \forceint
    = \nabla \cdot \sigmaint
    = \sum_{\alpha, \beta>\alpha}^{N} \sigma_{\alpha\beta}
        \left[I_{\alpha\beta}\kappa_{\alpha\beta} \vec{n}_{\alpha\beta} +
        \mat{P} \cdot \nabla I_{\alpha\beta} \right]
\label{eq:force_int}
\end{dmath}
where $\kappa_{\alpha\beta} \equiv \nabla \cdot \vec n_{\alpha\beta}$ is the curvature of the interface between phases $\alpha$ and $\beta$.
Note that in Eq.~(\ref{eq:force_int}), $\forceint$ is a vector and stands for force per unit volume acting on the interface and should not be confused with the free energy density $f$ which is a scalar quantity.
In Eq.~(\ref{eq:force_int}), the contribution to $\forceint$ arising from $\nabla I_{\alpha\beta}$ is projected onto the tangential plane.
Therefore, the second term in this equation vanishes at all interfaces, where only two phases meet.
This results from the fact that, in this case, $\phi_\alpha+\phi_\beta=1$ and thus $\nabla I_{\alpha\beta}$ is parallel to $ \vec n_{\alpha\beta}$.

\subsection{Simulation details}

We use an iterative algorithm~\cite{hu_phase-field_2001}, based on Fourier transformation, to solve Eq.~(\ref{eq:mech-equil}).
A boundary condition is used which allows a free volume expansion (see~\cite{steinbach_multi_2006}).
The three dimensional space is discretized by equally spaced lattice nodes.
As mentioned above, the stress induced by interface energy becomes relevant for elastic deformation of solid bodies on the nanometer scale.
Therefore, the lattice spacing has been set to $\Delta x = 10^{-9}$m.
If not mentioned otherwise, the interface width is set to $\eta = 10 \Delta x$.
For the surface or interface energy, we set $\intenergy_{\alpha\beta} = 1$J/m$^2$, which gives the correct order of magnitude for surface energy of metals (see, e.g.~\cite{maiya_surface_1967} for surface energy of Nickel).

\section{Results and discussion}
\label{sec:results}

We have seen above that the pressure increase inside a liquid drop is determined by the curvature of its surface (see Eq.~(\ref{eq:young_laplace})).
For the case of a homogeneous and isotropic solid, the stiffness tensor is uniquely determined by only two independent elastic constants, conveniently written as $C_{ijkl} =\lambda \delta_{ij} \delta_{kl} + \mu \left(\delta_{ik} \delta_{jl} + \delta_{il} \delta_{lj} \delta_{jk} \right)$ where $\lambda$ and $\mu$ are the first and second Lam\'{e} parameters, respectively.
Transferring Eq.~(\ref{eq:mech-equil}) into spherical coordinates and using the isotropic stiffness tensor, one can analytically solve a number of interesting problems which can then serve as telling benchmark of the present methodology.
Examples investigated here include spherical heterogeneities~\cite{sharma_effect_2003} such as an elastic body embedded in a highly compliant medium and a spherical inclusion (gas or solid) in a hard elastic matrix as well as a thin planar sheet made of an elastic material.

In the following, a detailed validation of the proposed model is provided via these analytically solvable problems.
The model is then used to study the stress distribution within more complex bodies such as an ellipsoid and a sintered structure revealing heterogeneous stress distribution due to a spatially variable curvature.

\subsection{Spherical inhomogeneities}
\label{sec:spherical_bodies}

Consider an isotropic elastic matrix $\alpha$ with a spherical isotropic elastic inclusion $\beta$ with the radius $R$ and the interface energy $\sigma_{\alpha\beta}$.
The pressure difference across the interface is then given by (see App.~\ref{app:inhomogeneities} for a derivation)
\begin{dmath}
    \label{eqn:delta_p}
    \Delta p =
    \frac{3 \lasphere + 2 \musphere}{4 \muouter + 3 \lasphere + 2 \musphere}
    \frac{2 \sigma_{\alpha\beta}}{R}
\end{dmath},
where $\lasphere$ and $\musphere$ are the Lam\'e parameters of the inclusion and
respectively $\laouter$ and $\muouter$ the Lam\'e parameters of the matrix.
In the following we consider also the radial component $\sigma_{rr}$ of the stress tensor, given by
\begin{dmath}
    \label{eqn:sigma_rr}
    \sigma_{rr}
    =
    \begin{cases}
        - \frac{3 \lasphere + 2 \musphere}{4 \muouter + 3 \lasphere + 2 \musphere} \frac{2 \sigma_{\alpha\beta}}{R}& r < R \\
        + \frac{4\muouter}{4 \muouter + 3 \lasphere + 2 \musphere} \frac{2 \sigma_{\alpha\beta} R^2}{r^3} & r > R
    \end{cases}
\end{dmath}.
Other components of the stress tensor are given in Eq.~(\ref{eqn:inhomo_stress}).

As a first example, a spherical elastic body surrounded by a `gas' phase shall be investigated here.
The `gas' phase is modeled as an extremely compliant medium.
For the elastic constants of the solid, we choose $\lasphere = \musphere = 100 \, \text{GPa}=10^{11}\text{Pa}$.
Recalling the value of the interface energy $\sigma_{\alpha\beta}=1$J/m$^2$, the Laplace pressure $\sigma_{\alpha\beta}\kappa$ in the present study is of the order of $0.1 \text{GPa}$ for a solid sphere of radius $R=20$nm.
Thus, the assumption of linear elasticity $\left\Vert \mat \varepsilon \right\Vert \ll 1$ is fulfilled in this work, where the typical linear sizes investigated are roughly a few tens of nanometer.

\begin{figure}
    \subfloat[]{\includestandalone[width= \columnwidth,mode=buildnew]{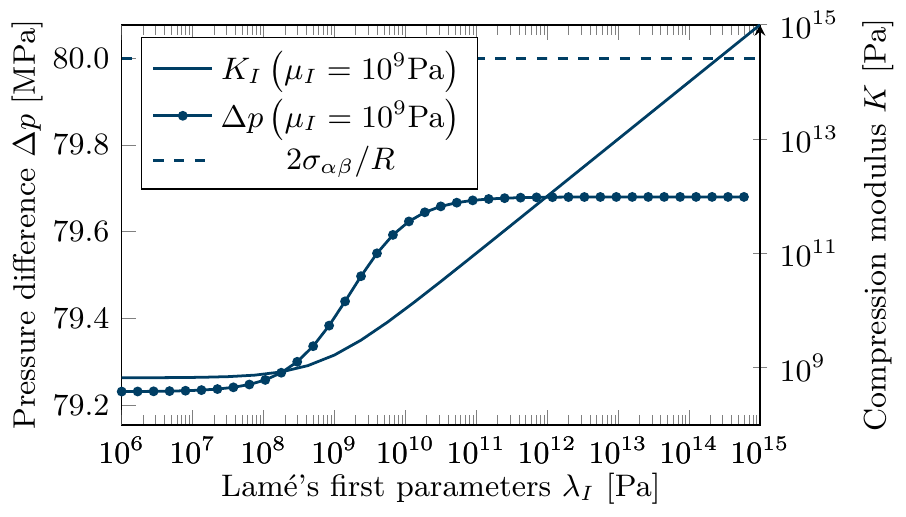}}\\
    \subfloat[]{\includestandalone[width= \columnwidth,mode=buildnew]{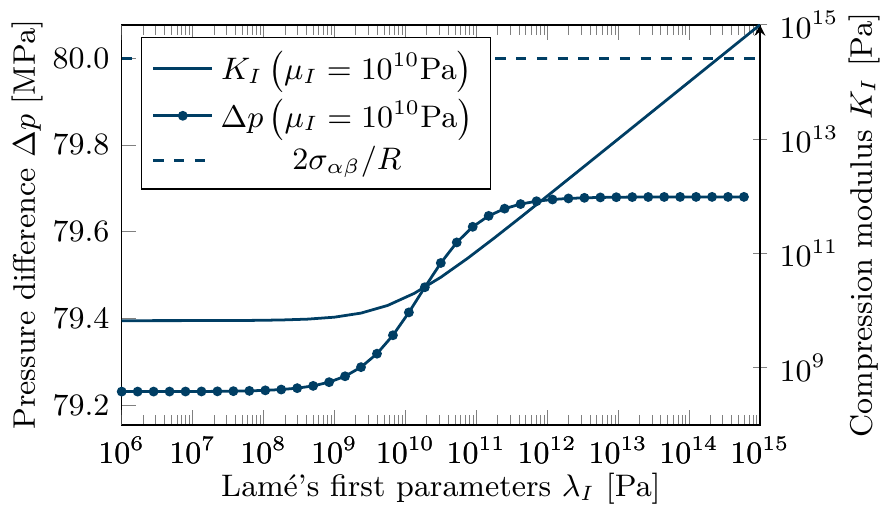}}\\
    \subfloat[]{\includestandalone[width= \columnwidth,mode=buildnew]{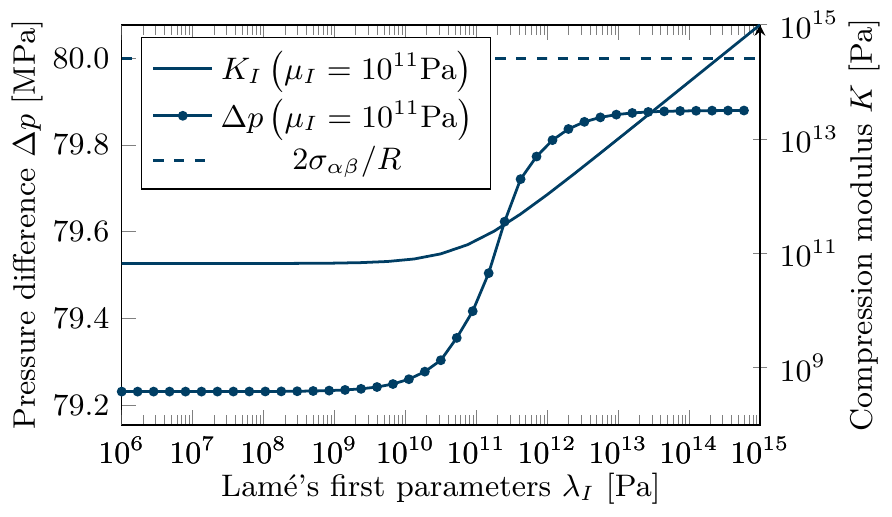}}
    \caption[]{Dependence of the pressure difference $\Delta p$ between an elastic spherical body and the surrounding `gas' phase on its elastic properties.
        The radius of the sphere is $R=25$nm.
        The surrounding medium is modeled as a highly compliant material with Lam{\'e} parameters $\laouter=\muouter=10^{-2}\Pa$.
        In each plot, $\Delta p$ is depicted versus the Lam{\'e} first parameter $\lasphere$, while the second Lam{\'e} parameter $\musphere$, is kept constant at the values of (a) $\musphere=10^{11}\Pa$ (b) $\musphere=10^{10}\Pa$ and (c) $\musphere=10^{9}\Pa$.
        The right vertical axis is used to survey the variation of the compression modulus $K_{\text{\scriptsize{I}}}=\lasphere+2\musphere/3$.
        All computations are performed with an interface width of $\eta = 6 \Delta x$.}
    \label{fig:stiffness_dependency}
\end{figure}

For the `gas' phase, we set $\laouter = \muouter = 10^{-2}$Pa.
\label{page:lambdagas} It is important to realize the large ratio of the stiffness constants between the two phases,
$\left \Vert \mat C_{\text{solid}} \right \Vert / \left \Vert \mat C_{\text{gas}} \right \Vert = 10^{13}$ with the L2-Norm $\left \Vert \mat C \right \Vert = \sqrt{\sum_{ijkl} C_{ijkl}^2}$.

\subsubsection[Curvature dependency]{Pressure inside an elastic body surrounded by a gas}
\label{sec:results_sphere_young}

As a first result, the pressure difference $\Delta p$ between inner and outer parts of a spherical elastic body embedded in a gas with $\laouter = \muouter = 10^{-2}$Pa is shown in Fig.~\ref{fig:stiffness_dependency} for different choices of $\lasphere$ and $\musphere$.
The pressure is calculated as the one third of the trace of the stress tensor, $p = - \frac{1}{3} \operatorname{trace}\left( \mat \sigma \right)$.
Despite the variation of the first Lam\'e parameter $\lasphere$ by roughly 7 decades, the pressure difference is essentially constant, varying by approximately $0.5\%$ only.

In order to highlight the variation of elastic properties of the sphere upon a change of $\lasphere$, the compressibility is also shown in Fig.~\ref{fig:stiffness_dependency}.
Depending on the value of $\musphere$, it varies over three ($\musphere=10^{11}\Pa$) to six ($\musphere=10^{9}\Pa$) decades.
Using the expression for the Poison ratio, $\nu=\lambda/(2(\lambda+\mu))$, it is also seen that, in all the three investigated cases, the Poisson ratio of the sphere varies between roughly 0 ($\lasphere \to 10^{6}\Pa \ll \musphere$) and 1/2 ($\lasphere \to 10^{15}\Pa \gg \musphere$).

The main reason behind the insensitivity of $\Delta p$ with respect to a variation of $\lasphere$ in Fig.~\ref{fig:stiffness_dependency} is that the elastic constants of the surrounding medium are negligibly small as compared to that of the sphere.
Indeed, values of $\lasphere$ and $\musphere$ associated with the sphere are at least by eight decades larger than those of the surrounding medium in the entire investigated range.
A survey of Eq.~(\ref{eqn:delta_p}) reveals that, in this case, the classical Young-Laplace equation, which describes the pressure difference between a liquid drop and the surrounding vapor (Eq.~(\ref{eq:young_laplace})) becomes an excellent approximation.

\begin{figure*}
    \subfloat[\label{fig:pressure_radius}]
    {\includestandalone[width=\columnwidth-2mm, mode=buildnew]{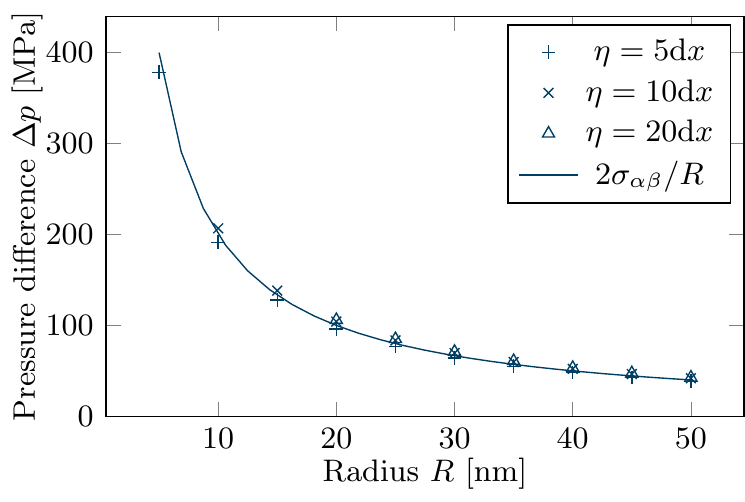}}
    \hspace*{4mm}
    \subfloat[\label{fig:sphere_stress}] {\includestandalone[width=\columnwidth-2mm, mode=buildnew]{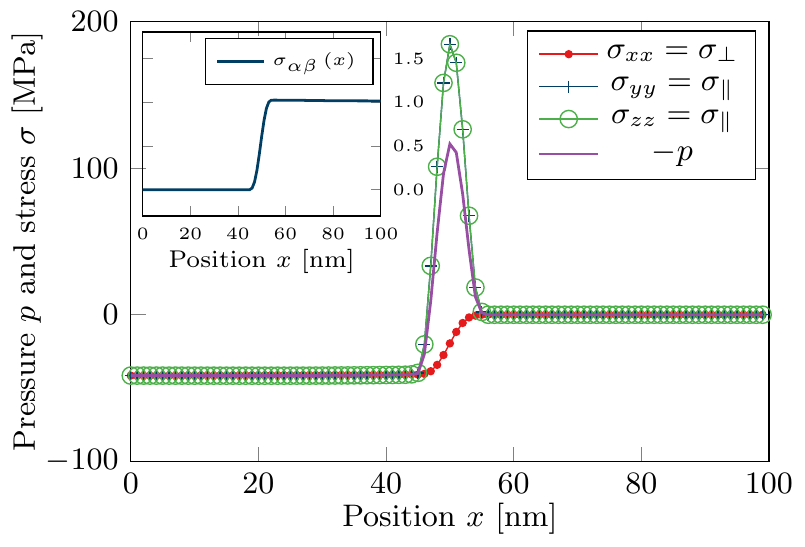}}
    \caption{(a) The dependency of the pressure difference $\Delta p$ between the inside of an elastic spherical body and the surrounding `gas' phase, on the radius of sphere $R$.
    The `gas' phase is represented here as an extremely soft material with Lam\'e parameters $\laouter = \muouter = 10^{-2} \text{Pa}$, while the sphere is an elastic body with $\lasphere = \musphere = 100 \text{GPa}= 10^{11} \text{Pa}$.
    Symbols correspond to different choices of the interface thickness $\eta$ as indicated.
    The solid line gives the analytic result, Eq.~(\ref{eqn:delta_p}), which reduces to the well-known Young-Laplace law for the present set of Lam\'e parameters.
(b) The magnitude of pressure $p = -\frac{1}{3} \text{trace} \left(\mat \sigma \right)$ and components of the stress tensor $\mat \sigma$ alongside the $x$ axis for a spherical body of Radius $R=50$nm surrounded by a `gas' phase.
    The Lam\'e parameters are identical to (a).
    The center of the sphere is at the coordinate origin.
    Note that the diagonal components of the stress tensor parallel to the interface are equal: $\sigma_{yy} = \sigma_{zz} = \sigma_{\parallel}$.
The normal component $\sigma_{xx} = \sigma_{\bot}$, however, behaves differently within the interface.
This difference, when integrated along the interface normal, is the surface energy $\intenergy_{\alpha\beta}=\lim_{x\to \infty}\intenergy_{\alpha\beta} (x)$, where $\intenergy_{\alpha\beta} (x)= \int_0^x \diff n \left(\sigma_\parallel - \sigma_\bot \right)$.
The inset shows this integral as a function of the integration limit $x$.
As expected, it approximately reaches the value of $\intenergy_{\alpha\beta} = 1$ N/m when passing through the interface.
}
        \label{fig:solid-sphere-in-gas}
\end{figure*}

This fact is further highlighted in Fig.~\ref{fig:pressure_radius}, where the pressure difference $\Delta p$ is shown as function of the sphere's radius for three different choices of the interface thicknesses, $\eta=5\Delta x$, $10\Delta x$ and $20\Delta x$.
Obviously, the data on $\Delta p$ are in very good agreement with Eq.~(\ref{eqn:delta_p}) for all the three interface thicknesses investigated.

To illustrate that the present multiphase-field approach also provides detailed information on spatial variation of the stress tensor across the interface, Fig~\ref{fig:sphere_stress} shows the pressure $p$, the normal $\sigma_{\bot}$ and the tangential $\sigma_{\parallel}$ components of the stress tensor along a center line of a spherical solid body surrounded by a `gas' phase.
All three quantities are zero outside the sphere, but approach a finite and spatially constant value close to its center.
Within the interface domain, however, strong variations are observed.
These variations of $\sigma_\bot$ and $\sigma_\parallel$ contain important information about the surface or interface energy.
Indeed, the mechanical definition of the surface or interface energy between two phases $\alpha$ and $\beta$ reads~\cite{rowlinson_molecular_1982,varnik_molecular_2000}
\begin{dmath} \label{eq:DefSurfaceTension2}
    \intenergy_{\alpha\beta} = \int_0^\infty \diff n
    \left(\sigma_\parallel - \sigma_\bot \right),
\end{dmath}
where the integral is performed along the direction normal to the surface or interface.

Equation~(\ref{eq:DefSurfaceTension2}) allows to easily differentiate between the energy, which is necessary to create a
surface of unit area $\intenergy_{\alpha\beta}$ and the tangential stress $ \mat \sigma_\parallel $ acting on the surface.
Moreover, it also provides a simple way to test Eqs.~(\ref{eq:mech-equil}) and (\ref{eq:sigmatotal}): Since the surface or interface energy $\sigma_{\alpha\beta}$ is an input parameter and thus known from the outset, it can be checked whether the stress tensor obtained from the solution of these equations satisfies Eq.~(\ref{eq:DefSurfaceTension2}).
The result of such a test is shown in Fig.~\ref{fig:sphere_stress}, where the integral in Eq.~(\ref{eq:DefSurfaceTension2}) is evaluated.
The performed calculation delivers the correct surface energy of $\intenergy_{\alpha\beta} = 1$J/m$^2$ to a good approximation.

\subsubsection[Material Dependency]{Pressure inside spherical bodies surrounded by a solid matrix}
\label{subsec:results_sphere_material}

\begin{figure*}
    \subfloat[\label{fig:pressure_2}]{\includestandalone[width=\columnwidth-2mm, mode=buildnew]{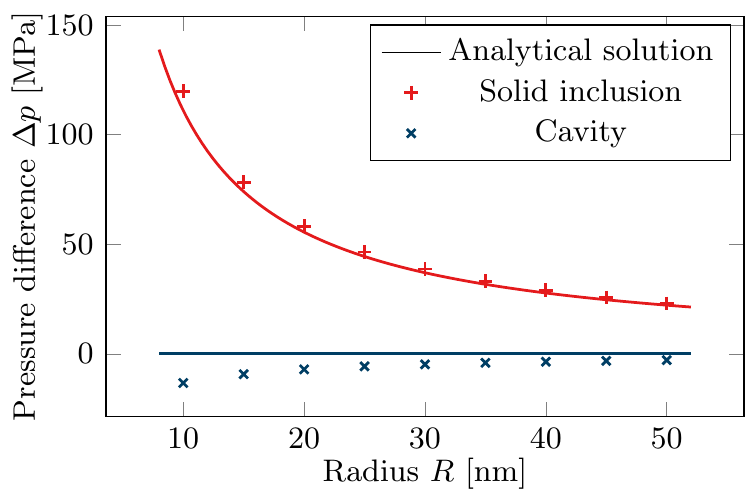}}
    \hspace*{4mm}
    \subfloat[\label{fig:sphere_stress_2}]{\includestandalone[width=\columnwidth-2mm, mode=buildnew]{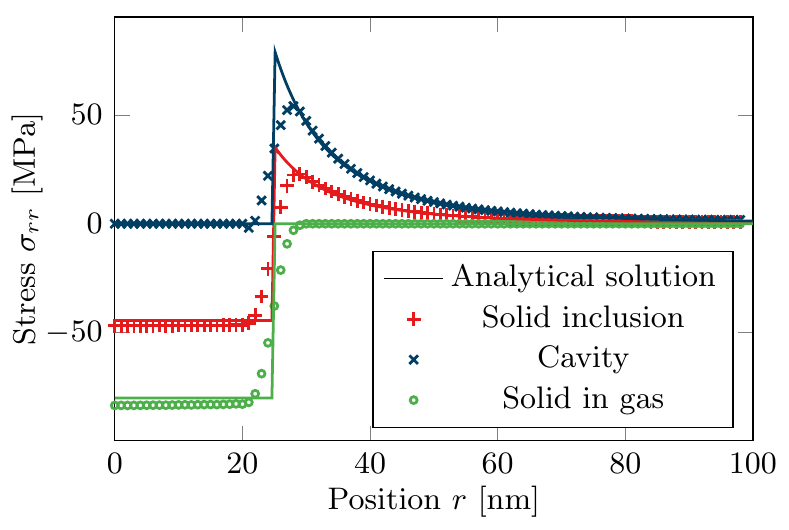}}
    \caption{(a) Pressure difference $\Delta p$ between a spherical inclusion (gas or solid) and the surrounding solid matrix versus the radius $R$ of inclusion.
    The solid inclusion has the same Lam\'e parameters as the elastic matrix, $\laouter = \muouter = \lasphere=\musphere=100 \text{GPa}$.
    For the cavity, we chose $\lasphere = \musphere = 10^{-2} \text{Pa}$.
    Data obtained from the present multi-phase-field method are shown as symbols.
    The solid lines give the predictions of Eq.~(\ref{eqn:delta_p}), which, for the present cases of a solid and gas inclusion, yields $\Delta p =10\intenergyab/(9R)$ and $\Delta p=0$, respectively.
    The observed deviations in the case of a cavity are very probably due to the fact that the radius of sphere becomes comparable to the interface width $\eta=10\Delta x=10$nm.
       (b) The radial component of the stress tensor $\sigma_{rr}$ is shown versus the distance from the sphere's center for the three test cases of (1) a spherical solid inclusion in a solid matrix and (2) a spherical cavity in a solid body and (3) a solid sphere in a gas environment.
       Symbols show results obtained within the present model and solid lines give the predictions of Eq.~(\ref{eqn:sigma_rr}) for the respective cases.
       }
       \label{fig:pressure_radius_non_hydro}
\end{figure*}

In the above, we considered a homogeneous isotropic solid sphere surrounded by a highly compliant medium (a `gas') and tested the present model against the corresponding analytical solution, Eq.~(\ref{eqn:delta_p}).
Here, we investigate a new situation where an elastic solid matrix with $\laouter = \muouter = 100 \text{GPa}$ contains a spherical inclusion made of either a solid material (referred to a solid inclusion) or gas (cavity).
In the first case, we set $\lasphere = \musphere = \laouter = \muouter=100 \text{GPa}$ and in the second case we chose $\lasphere = \musphere = 10^{-2} \text{Pa}$.

Simulation results on this issue are depicted in Fig.~\ref{fig:pressure_radius_non_hydro} and are compared to the predictions of Eq.~(\ref{eqn:delta_p}).
For the two cases considered here this equation predicts $\Delta p =\frac{9\intenergyab}{10R}$ for solid inclusion and $\Delta p =10^{-13}\times\frac{5\intenergyab}{2R} \approx 0$ for cavity, respectively.
The observed deviations between simulation results and the analytic predictions can be rationalized by recalling that Eq.~(\ref{eqn:delta_p}) is obtained by assuming an infinitely sharp interface between the adjacent phases.
The present model, on the other hand, introduces a finite interface width in order to smooth out discontinuities.
As the radius of sphere becomes comparable to this thickness, deviations from a sharp interface solution are to be expected.
This interpretation is in line with the fact that the simulation results approach the analytic prediction for large $R$.

As anther telling test with a focus on spatially varying stress field,
Fig.~\ref{fig:sphere_stress_2} shows the radial component of the stress tensor for the above investigated cases.
One can see that the numerically obtained results agree well with the analytical solution Eq.~(\ref{eqn:sigma_rr}) outside the interface domain.
It is noteworthy that there is no sharp interface-analog of the stress tensor variations within the interface domain.
Nevertheless, we emphasize here that the behavior of stress tensor within the interface is consistent with other physical properties of the model (see the inset of Fig.~(\ref{fig:sphere_stress_2})).

In order to better understand the results of the cavity, a detailed analysis of the forces acting in the interface domain proves to be very useful.
For this purpose, we first decompose the force density $\force$ into one part associated with the interface and another part assigned to the bulk:
\begin{dmath}\label{eq:froce_split}
    \force = \forceint + \forcebulk
\end{dmath}.
The former has been introduced in Eq.~(\ref{eq:force_int}).
For the latter, we have
\begin{dmath}\label{eq:froce_bulk}
    \forcebulk
    = \nabla \cdot \sigmabulk
\end{dmath}.

\begin{figure*}
    \subfloat[\label{fig:force_balance_pressure}] {\includestandalone[width=\columnwidth-2mm, mode=buildnew]{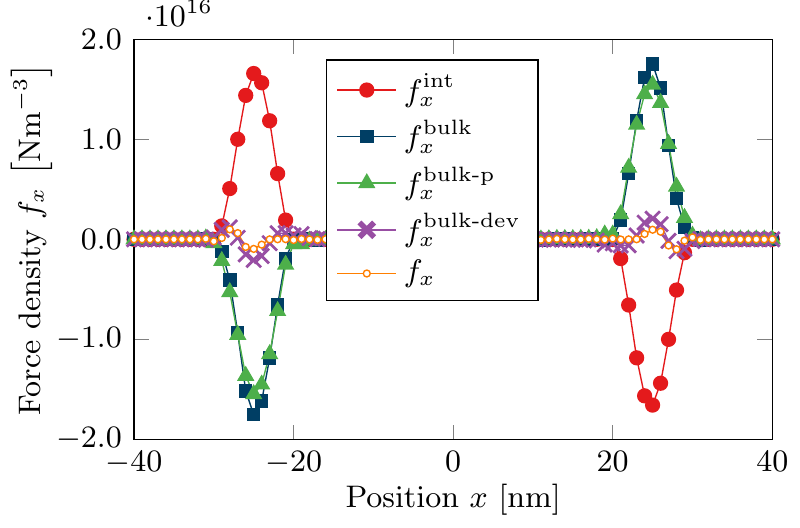}}
    \hspace*{4mm}
    \subfloat[\label{fig:force_balance_stress}] {\includestandalone[width=\columnwidth-2mm, mode=buildnew]{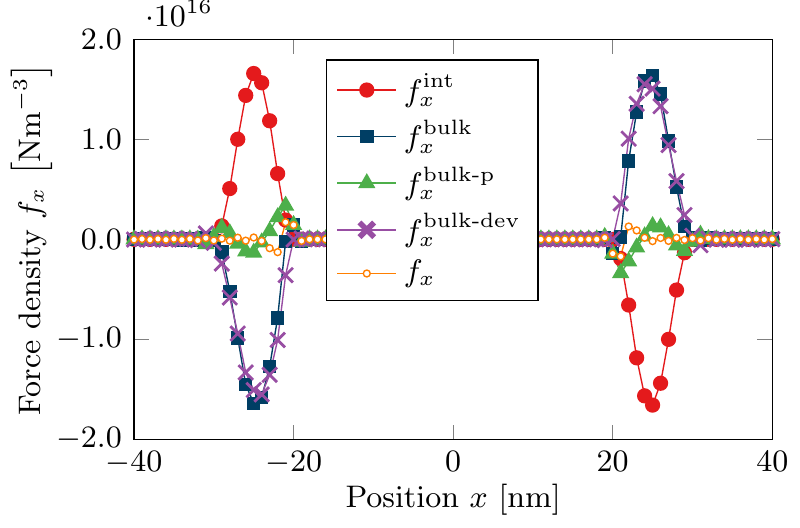}}
    \caption[]{The $x$ component of the force densities across the center line of a spherical solid body along the $x$ axis with $\force = \forceint + \forcebulk$ and $\forcebulk = \forcepressure + \forcedeviatoric$ (see Eqs.~\ref{eq:force_int},~\ref{eq:froce_split},~\ref{eq:froce_bulk}, and~\ref{eq:froce_bulk_split}).
    The total force density is split into a part arising from the interface energy $\forceint$ and a part related to deformation in the bulk.
    These two contributions compensate each other at mechanical equilibrium.
    (a) A solid sphere surrounded by gas ($\left \Vert \mat C_{\text{outer}} \right \Vert \ll \left \Vert \mat C_{\text{inner}} \right \Vert$).
    The contribution to $\forcebulk$, which compensates $\forceint$ originates from pressure gradient.
    (b) A gas cavity within a solid matrix ($\left \Vert \mat C_{\text{outer}} \right \Vert \gg \left \Vert \mat C_{\text{inner}} \right \Vert$).
    In this case, this is the divergence of the deviatoric stress tensor $\forcedeviatoric$ which mainly contributes to $\forcebulk$.}
    \label{fig:force_balance}
\end{figure*}

Figure~\ref{fig:force_balance} shows the above defined force densities across the center line of a spherical solid body along the $x$ axis.
One can see that all force densities are non-zero only in the interface and that $\forceint$ and $\forcebulk$ essentially compensate each other.
In the both cases shown in Fig.~\ref{fig:force_balance_pressure} and Fig.~\ref{fig:force_balance_stress}, although not entirely zero, the remaining total force density $\force$ is relatively small compared to the magnitude of $\forceint$ and $\forcebulk$.
The non-zero value of $\force$ within the interface is presumably due to numerical discretization errors.
Indeed, as addressed in a recent study~\cite{vakili_numerical_2017}, the accurate representation of forces within diffuse interfaces is a challenging numerical task and requires sophisticated stencils.
The implementation of a higher order stencil would certainly increase the numerical accuracy and shall be considered if the information within the interface domain is of major interest.

It is instructive to split $\sigmabulk$ into contributions arising from hydrostatic pressure and deviatoric stress $\sigmabulk = \mat s + \mat 1p$.
Recalling Eq.~(\ref{eq:froce_bulk}), this leads to a natural splitting of $\forcebulk$ into
\begin{dmath}
    \forcebulk = \forcepressure + \forcedeviatoric,
    \label{eq:froce_bulk_split}
\end{dmath}
with
\begin{dgroup}
\begin{dmath}
\forcepressure = \nabla p\;\;\;\text{and}
\label{eq:force_bulk-p}
\end{dmath}
\begin{dmath}
\forcedeviatoric = \nabla \cdot \mat{s}.
\label{eq:force_bulk-dev}
\end{dmath}
\end{dgroup}

These quantities are also shown in Fig.~\ref{fig:force_balance}.
Interestingly, in the case of a solid sphere in a gas environment, the main contribution which compensates the surface energy-induced force arises from the gradient of hydrostatic pressure $\forcepressure$ (Fig.~\ref{fig:force_balance_pressure}).
In the opposite situation of a cavity inside a solid matrix, on the other hand, this is the divergence of the deviatoric stress $\forcedeviatoric$ which ensures force balance (Fig.~\ref{fig:force_balance_stress}).
Apparently, the significant increase of deviatoric stress is due to non-negligible elasticity of the matrix surrounding the sphere.
As the sphere contracts in response to the action of surface-induced radial forces, the surrounding matrix must expand accordingly, giving rise to elastic restoring forces which act opposite to the direction of contraction.
Since the matrix is not a sphere but a cube with a spherical hole, its deformation is quite complex with non-zero off-diagonal components.
However, if the matrix is not a solid body but a `gas'-like medium, its resistance to the contraction of the sphere is negligible so that the only way to compensate the surface energy-induced force is via the increase of internal pressure.

\subsection{Stress within non-spherical solids}
\label{sec:non_spherical_solids}
As further applications of the proposed multi-phase-field method for surface energy-induced deformation, we consider here a thin solid plate, an elliptical body and the more complex geometry of two solid spheres after having undergone a sintering process.

\subsubsection{Stress difference without curvature}
\label{sec:plate}
\begin{figure*}
    \subfloat[\label{fig:plate_stress}]
        {\includestandalone[width=\columnwidth-2mm,mode=buildnew]{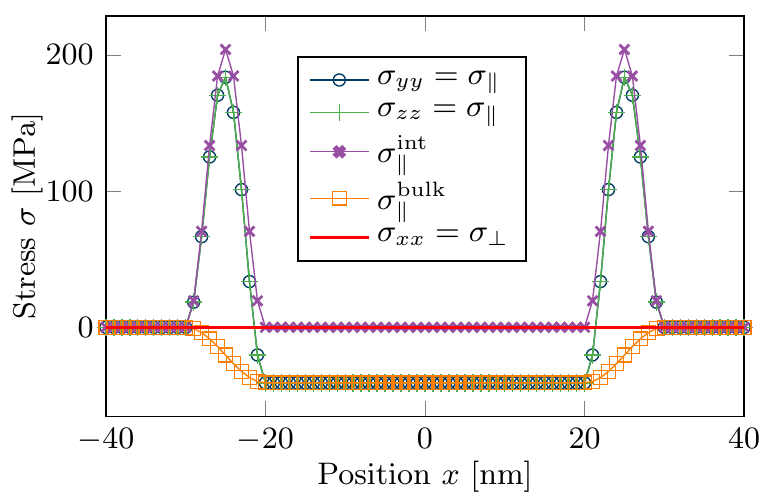}}
    \hspace*{4mm}
    \subfloat[\label{fig:plate_pressure_diameter}]
        {\includestandalone[width=\columnwidth-2mm,mode=buildnew]{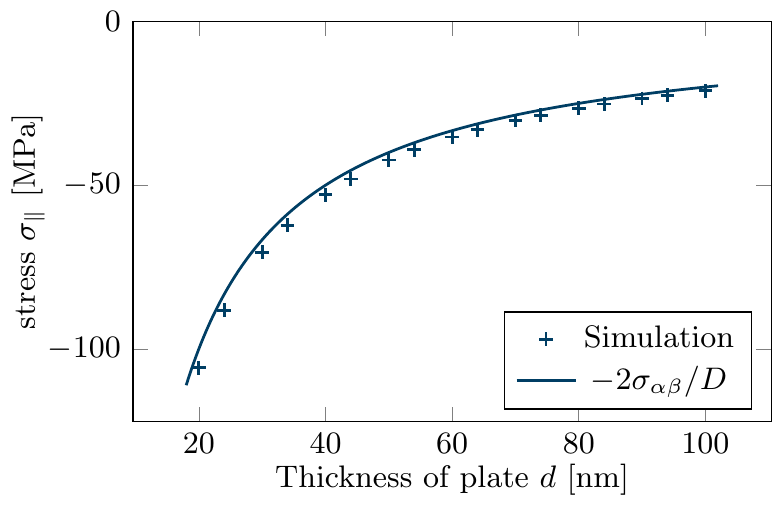}}
    \caption[]{(a) The tangential components of the stress tensor $\sigma_{yy}$ and $\sigma_{zz}$ across the surface normal direction of a solid plate of thickness $D=50$nm surrounded by a `gas' phase.
        The interface width is set to $\eta = 10 \Delta x=10$nm.
        As expected, the components of stress tensor along the two tangential directions $y$ and $z$ are equal across the interface: $\sigma_{yy}=\sigma_{zz}$.
        The normal component $\sigma_{xx}$ is zero since no force is acting along the direction normal to the interface.
        Also, there is no force acting parallel to the surface, because $\sigma_{\parallel}$ is constant alongside the interface, so that $ \nabla \cdot \sigma = \sum_i \partial_i \sigma_{ij} \vec{e}_j = \vec{0}$ (note that all the off-diagonal components of the stress tensor are zero).
        The plot also shows the interface and bulk contributions to $\sigma_{\parallel}=\sigma_{\parallel}^{\text{int}}+\sigma_{\parallel}^{\text{bulk}}$.
        (b) Simulation results on the tangential stress within the plate $\sigma_{\parallel}$ are shown versus the plate thickness $D$.
        The solid line is the analytic prediction given by Eq.~(\ref{eq:sigma-plate}).}
    \label{eq:plate}
\end{figure*}

Due to the presence of curvature in a spherical body, the resulting stress and deformation appears as a natural consequence of the action of surface tension.
We show here that, interestingly, surface energy-induced deformation may also occur in the case of a planar surface or interface, i.e., in the absence of curvature.
Assuming that the stress tensor entirely vanishes outside the plate, the condition of mechanical equilibrium then implies that the component of the stress tensor in the direction normal to the surface is zero.
Tangential stresses, however, are present and lead to a deformation in the tangential direction.
By minimizing the free energy (see.
App.~\ref{app:plate}), one obtains
\begin{dmath}\label{eq:sigma-plate}
    \sigma_{\parallel} = -\frac{2\sigma_{\alpha\beta}}{D}
\end{dmath},
where $\sigma_{\parallel}$ is the component of the stress tensor parallel to the surface and $D$ is the thickness of the plate.
Figure~\ref{eq:plate} shows a good agreement between the results obtained from simulations using the present model and the analytic prediction given in Eq.~(\ref{eq:sigma-plate}).
The plot also illustrates the spatial variation of the tangential components of the stress tensor and the corresponding contributions of the bulk and interface stresses.

\subsubsection{An ellipsoidal elastic body}
\label{sec:ellipsoidal_bodies}

\begin{figure}
    \subfloat[\label{fig:ellipsoid_pressure}]
        {\includegraphics[width=\columnwidth,height=0.65\columnwidth]{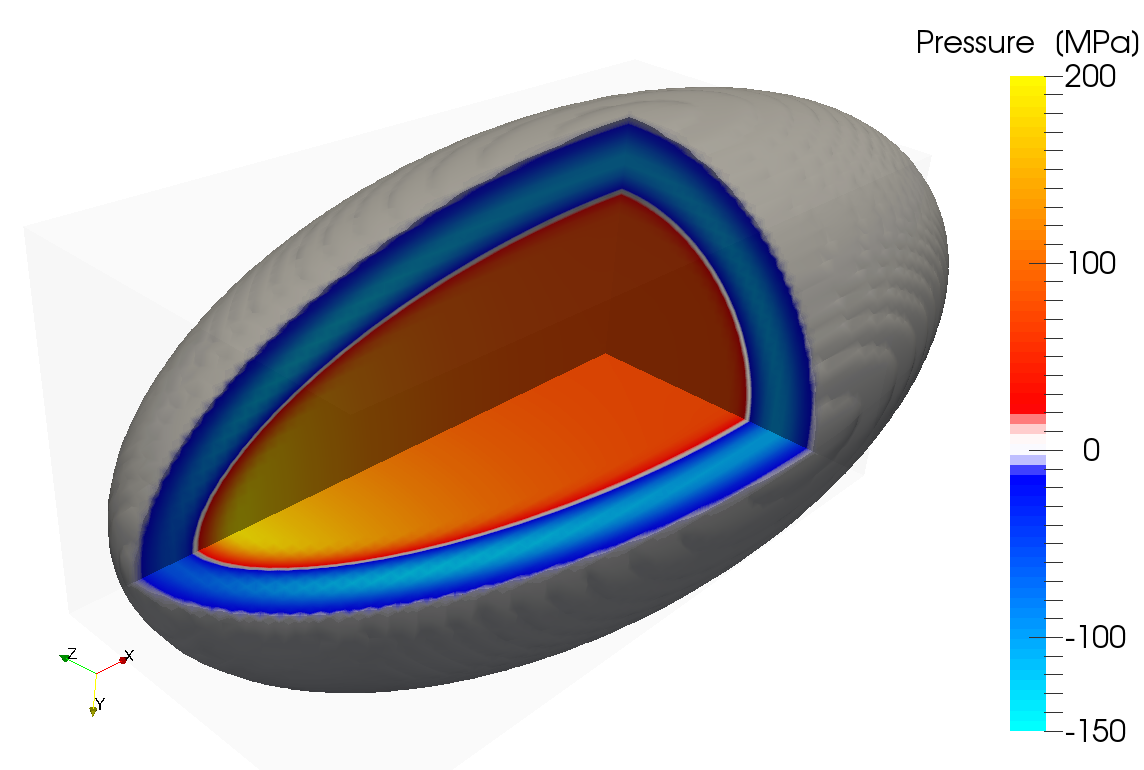}}\\

    \subfloat[\label{fig:ellipsoid_stress_x}]
        {\includestandalone[width=\columnwidth, mode=buildnew]{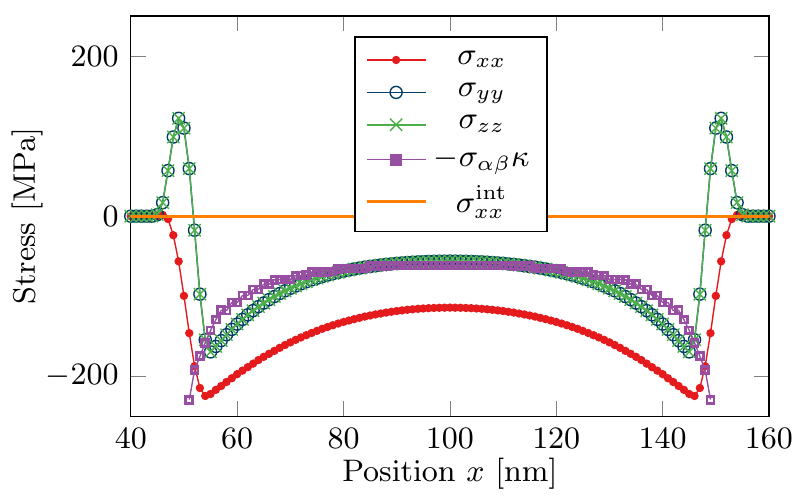}}\\

    \subfloat[\label{fig:ellipsoid_stress_z}]
        {\includestandalone[width=\columnwidth, mode=buildnew]{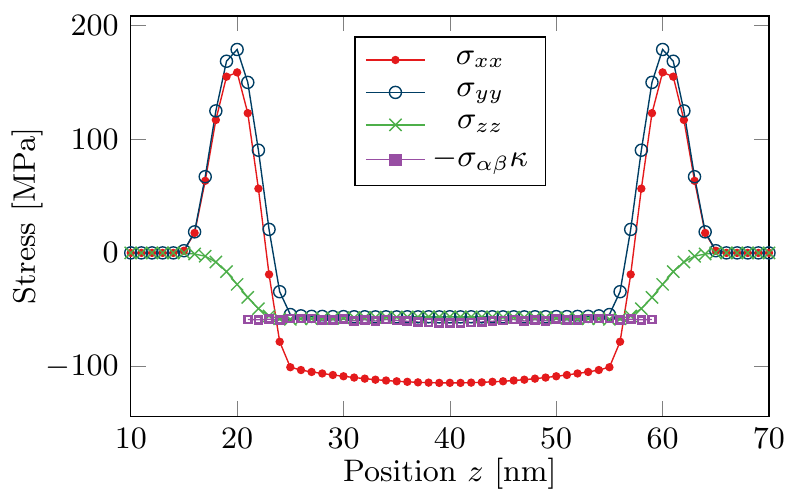}}

    \caption{(color online)(a) Color map of the pressure field inside an elastic ellipsoidal body.
        One can recognize the tension within the surface domain (blue) and the opposing hydrostatic pressure inside the bulk (yellow-orange).
        (b) Diagonal components of the stress tensor and the surface curvature term $-\intenergyab\kappa$ along the $x$ direction and through the line connecting both foci of the elliptical body and (c) along the $z$ direction and through the center of the elliptical body.
        The computation domain is discretized with $200 \times 100 \times 100$ lattice nodes.
        }
\end{figure}

Taking the example of an elliptical body, we investigate the pressure field inside the body at mechanical equilibrium.
Results of these simulations are illustrated in Fig.~\ref{fig:ellipsoid_pressure}.
It is important to note that, due to the loss of spherical symmetry, it is difficult to obtain an analytical solution for the corresponding equations of elasticity~\cite{sharma_size-dependent_2006}.
Nevertheless, a qualitative understanding of the simulation results is possible based on estimates of the relevant local curvature.
In order to better illustrate this aspect, the diagonal components of the stress tensor and the curvature term are compared in Fig.~\ref{fig:ellipsoid_stress_x} along the $x$ direction and through the line connecting both foci of the elliptical body.
In Fig.~\ref{fig:ellipsoid_stress_z}, the same quantities are shown along the $z$ direction through the center of the elliptical body.

One can see that the magnitude of $\sigma_{xx}$ at the sharp ends of the ellipsoid is comparable to $-\sigma_{\alpha\beta} \kappa$ at the same point in Fig.~\ref{fig:ellipsoid_stress_x}.
In the same way, the magnitude of $\sigma_{yy}$ and $\sigma_{zz}$ at the blunt sides of the ellipsoid is comparable to $-\sigma_{\alpha\beta} \kappa$ at the same point in Fig.~\ref{fig:ellipsoid_stress_z}.
The correlation can be understood by looking at Eq.~(\ref{eq:force_int}).
The interface-force density is proportional to the local curvature and the surface tension and acts along the surface normal direction.
Hence, at the sharp edges of the ellipsoid, $\vec{f}^{\text{int}}$ acts only along the $x$ direction and has to be compensated by a force of the bulk with $f^{\text{bulk}}_x = \partial_x \sigma^{\text{bulk}}_{xx}$, since there is no shear stress in the bulk.
This explains the gradient of the total stress $\sigma_{xx}$ at the sharp edges of the ellipsoid, which can be seen in Fig.~\ref{fig:ellipsoid_stress_x}.
Because there is no contribution of $\mat{\sigma}^{\text{int}}$ normal to the interface, $\sigma_{xx}$ and $\sigma^{\text{bulk}}_{xx}$ are equal in Fig.~\ref{fig:ellipsoid_stress_x} (see Eq.~(\ref{eq:sigmatotal}) and~(\ref{eq:sigmastar})).
Following a similar argument, the correlation of $\sigma_{xx}$, $\sigma_{yy}$ and the curvature term $-\sigma_{\alpha\beta} \kappa$, observed in Fig.~\ref{fig:ellipsoid_stress_z} can be rationalized.

\subsubsection{Pressure inside a sintered structure}
\label{sec:sintering}
In order to further highlight the above idea, we address the mechanical equilibrium condition for the case of two solid spheres, sintered via surface diffusion~\cite{schiedung_multi-phase-field_2017}.
The resulting sintered structure is used here as the starting point of the calculations of deformation and stress fields.

\begin{figure}
    \subfloat[\label{fig:pressure-sinter}] {\includegraphics[width=0.95\columnwidth,height=0.65\columnwidth]{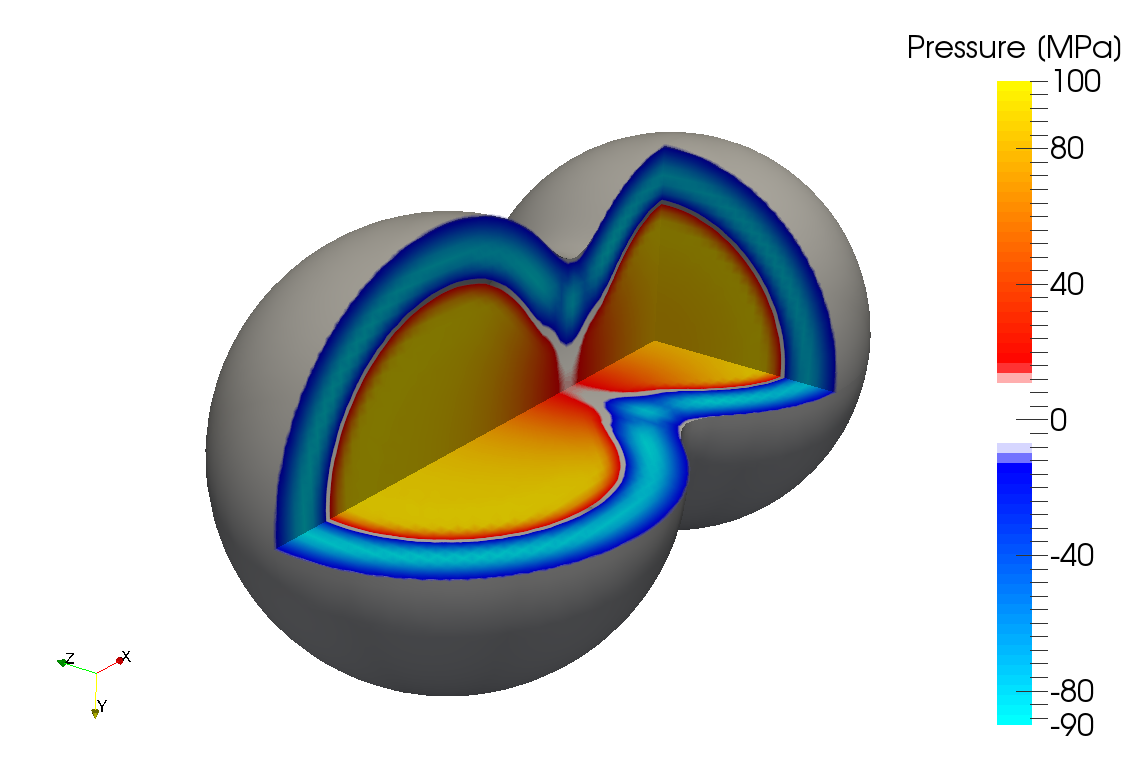}}

    \subfloat[\label{fig:pressure-sinter-line}] {\includestandalone[width=0.95\columnwidth, mode=buildnew]{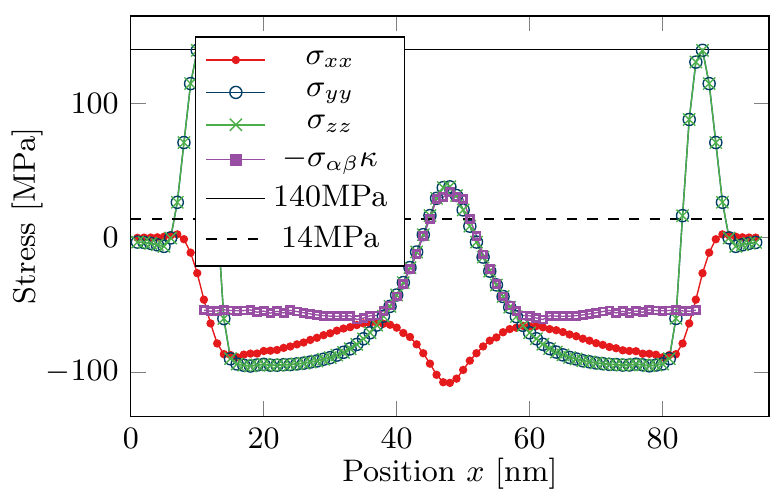}}
    \caption{(color online)(a) Color map of the pressure field $p = \frac{1}{3} \text{trace}\left( \mat \sigma \right)$ is shown alongside a cut into two spheres after having undergone a sinter process.
    (b) The diagonal components of the stress tensor $\sigma_{xx}$, $\sigma_{yy}$, $\sigma_{zz}$ and the curvature term $-\sigma_{\alpha\beta}\kappa$ are shown alongside the line connecting the centers of the two sintered spheres.
    The computation domain is discretized with $64\times 64\times 96$ lattice nodes.
    The surface and interface energies are set to $\intenergy_{\text{surface}} = 1$J/m$^2$ and $\intenergy_{\text{grain-boundary}} = 0.1$J/m$^2$.
        }
\end{figure}

Figure~\ref{fig:pressure-sinter} shows the thus obtained result for hydrostatic pressure alongside a cut into the two spheres after a sinter process and Figure~\ref{fig:pressure-sinter-line} shows the components of the stress tensor tangential and normal to the surface along the line connecting the centers of the two sintered spheres.
In these calculations, the interface energy of the grain boundary between the two spheres is assumed to be ten times smaller than the surface energy, $\sigma_{\text{grain-boundary}}=0.1$J/m$^2$.
As shown in Fig.~\ref{fig:pressure-sinter}, the pressure in the neck between the two grains is reduced compared to the pressure in the bulk of the two grains, but still positive.

Based on the initial setup, one would expect that the magnitude of tension in the grain-boundary is about ten percent of the magnitude of tension in the surface.
Indeed this can be seen by investigating the component of the stress tensor (see~\ref{fig:pressure-sinter-line}).
The amplitude of the tangential components of the stress tensor $\sigma_{yy}$ and $\sigma_{zz}$ is much larger in the surface as in the grain boundary in the middle of the computation domain.
Still, the magnitude of the resulting tensions in the interface is higher than expected.
This can be explained with the negative curvature of the neck which adds to the tension as well.
Figure~\ref{fig:pressure-sinter-line} also shows a comparison of the results with the curvature term $-\sigma_{\alpha\beta}\kappa$.
In the bulk of the grain, the curvature term does not deliver a valid description for the stress, but near the interface in the middle it corresponds to the magnitude of the tangential components of the stress tensor.

These results also demonstrate the capability of the present multi-phase-field method in studying complex geometries.

\section{Conclusion and outlook}
\label{sec:conclusion}
This work proposes a consistent treatment of surface energy and elastic deformation within the multi-phase-field framework.

The model is first validated using a number of benchmark problems for which analytic solution is available within the continuum elasticity theory.
These tests include various types of spherical inclusions in a gas or in a solid matrix.
Numerical results obtained with the present approach are found to be in good agreement with analytical predictions in all the investigated cases.
Effect of a finite interface thickness, inherent to all phase-field-type models, on the obtained results is also investigated.
It is shown that satisfactory data can be obtained already with an interface thickness of five or six times the lattice spacing.
In the case of a spherical cavity, deviations are observed in the limit of small cavities.
This is traced back to the fact that the length scale over which physical properties vary must be large compared to the interface width.

The present model also provides information about variation of stresses within the interface domain, for which no analog exists in the sharp interface equations of elasticity.
It is, nevertheless, known from molecular theories of capillarity that components of the stress tensor within the interface must obey certain conditions.
For example, the integral across the interface of the difference between the normal and tangential components of the pressure tensor is identical to the specific surface free energy.
Interestingly, the present model satisfies this fundamental requirement.

Moreover, the model is used to investigate the mechanical equilibrium conditions for a thin plate.
It is shown that a non-zero tangential stress within the solid plate occurs with a magnitude inversely proportional to the plate's thickness.
Simulation results are found to be in good agreement with the analytical result from theory of elasticity.

The strength of the method is demonstrated by the study of elastic deformation and the resulting stress distributions in cases for which, due to the complex geometry, analytic solutions are not available.
Two examples of an elliptical body and a sintered structure are chosen here for demonstration purpose.
Despite the lack of analytic prediction, a qualitative understanding of the stress distribution inside these bodies is gained by the use of local curvature as a key concept.

The proposed multi-phase-field method for surface energy-induced deformation can be easily combined with contributions to the free energy functional due to the action of electric and/or magnetic fields.
This would open the door to the application of the method to a wide range of physical phenomena on the nanoscale.

It is also noteworthy that, by considering the interface energy as a function of elastic deformation, it would be in principle possible to account for situations where an elastic enlargement of the surface or interface area reduces the free energy of the system.

\begin{acknowledgments}
    Financial support by the German Research Foundation DFG under the grand VA205/17-1 is gratefully acknowledged.
\end{acknowledgments}

\appendix

\section{Spherical inhomogeneities with surface tension}
\label{app:inhomogeneities}

In the following, we present analytical solutions for isotropic elastic bodies with spherical inhomogeneities and surface tension.
Because of the spherical symmetry of the considered cases, it is convenient to express the condition of mechanical equilibrium $\nabla \cdot \mat \sigma$ in spherical coordinates.
The equilibrium condition for the displacement field of an isotropic spherical body and a torsion-free body force $\vec{f}$ can be found in standard textbooks on elasticity (see, e.g. Landau and Lifshitz~\cite{landau_theory_1986}):
\begin{dmath}
    \label{eqn:eq_displacement_general}
    \left(\lambda + 2 \mu \right) \nabla \nabla \cdot \vec{u} = - \vec f
\end{dmath}.
Because of the symmetry of the problem and being torsion free, the displacement vector field can only have a radial dependency and be parallel to the vector $\vec r$: $\vec u \rightarrow u_r(r)\vec e_r$, where $\vec e_r=\vec r/r$ is the unit vector along the radial direction. This directly implies $\nabla \times \vec u = \vec 0$, which is the condition for being free of torsion. Thus, Eq.~(\ref{eqn:eq_displacement_general}) reduces to
\begin{dmath}
    \label{eqn:displacement_homo}
    \dfrac{\diff }{\diff r}
    \left[
       \dfrac{1}{r^2}
       \dfrac{\diff}{\diff r}
       \left(
           r^2 u_r
       \right)
    \right]
    =0
\end{dmath},
without a body force $\vec f$. Equation~(\ref{eqn:displacement_homo}) is solved by
\begin{dmath}
    u_r = C_1 r + \dfrac{C_2}{r^2}
    \label{eqn:displacement_solution}
\end{dmath},
with the two constants $C_1$ and $C_2$.
Equation~(\ref{eqn:displacement_solution}) describes the equilibrium displacement field of isotropic spherical symmetric bodies in the absence of body forces.
Although, forces may act on the surface of the body and enter through the boundary condition into Eq.~(\ref{eqn:displacement_solution}).

We follow a less general approach as in~\cite{sharma_effect_2003} and consider solely the effect of a constant surface tension without other forces.
Therefore, we consider a spherical inhomogeneity with the radius R, and the Lam\'e parameters $\lasphere$ and $\musphere$ surrounded by a matrix phase with the Lam\'e parameters $\laouter$ and $\muouter$.
By using Eq.~(\ref{eqn:displacement_solution}), we can write for the displacement field:
\begin{dmath}
    \label{eqn:displacement_ansatz}
    u_r =
    \begin{cases}
        C_1 r + \dfrac{C_2}{r^2} & r < R \\[3mm]
        C_3 r + \dfrac{C_4}{r^2} & r > R
    \end{cases}
\end{dmath}.
The displacement field should be finite inside the inhomogeneity, vanish for $r \rightarrow \infty$, and continuous at $r = R$ (Coherent interface) so that Eq.~(\ref{eqn:displacement_ansatz}) can be reduced to
\begin{dmath}
    \label{eqn:displacement_ansatz_2}
    u_r =
    \begin{cases}
        C r & r < R \\
        \dfrac{CR^3}{r^2} & r > R
    \end{cases}
\end{dmath}.
The remaining constant $C$ can by determined by the requirement of force balance at the interface
\begin{dmath}
    \label{eqn:force_balance}
    \left.
    \nabla \cdot \mat \sigma^{\text{bulk}}\right\vert_{r=R} + \left.
    \nabla \cdot \mat \sigma^{\text{int}} \right\vert_{r=R} = 0
\end{dmath}.
Here (see App.~\ref{app:proof}), the requirement of force balance simplifies to
\begin{dmath}
    \label{eqn:jump_condition}
    \lim_{r\rightarrow R^{+}} \sigma^{\text{bulk}}_{rr} -
    \lim_{r\rightarrow R^{-}} \sigma^{\text{bulk}}_{rr} = \frac{2\sigma}{R}
\end{dmath},
where $\lim_{r\rightarrow R^{+}}$ and $\lim_{r\rightarrow R^{-}}$ are right and left hand-side limit.
In order to determine the remaining constant $C$, the resulting stress of the displacement given by Eq.~(\ref{eqn:displacement_ansatz_2}) has to be calculated first.
For isotropic elastic bodies, the relation of bulk stress $\mat \sigma^{\text{bulk}}$ and the strain $\mat \varepsilon$ can be formulated independent of the coordinate system:
\begin{dmath}
    \label{eqn:stress_strain}
    \mat \sigma^{\text{bulk}} = \lambda \operatorname{trace} \left(\mat\varepsilon\right)\mat I + 2 \mu \mat \varepsilon
\end{dmath},
where $\mat I$ is the identity tensor.
Since the displacement field, Eq.~(\ref{eqn:displacement_ansatz_2}), is known, the non zero component of the strain tensor field are given by
\begin{dmath}
    \label{eqn:strain_spherial}
    \mat \varepsilon
    = \frac{\partial u_{r}}{\partial r}\vec e_{rr}
    + \frac{u_{r}}{r} \left(\vec e_{\varphi\varphi} + \vec e_{\theta\theta}\right)
\end{dmath},
where $\vec e_r$, $\vec e_\varphi$ and $\vec e_\theta$ are the basis vectors in spherical coordinates with the short hand of the dyadic product $\mat e_{\varphi\varphi} = \vec e_\varphi \vec e_\varphi$.
Inserting Eq.~(\ref{eqn:displacement_ansatz_2}) into Eq.~(\ref{eqn:strain_spherial}) and Eq.~(\ref{eqn:strain_spherial}) into Eq.~(\ref{eqn:stress_strain}) delivers for the stress tensor
\begin{dmath}
    \mat \sigma^{\text{bulk}}
    =
    \begin{cases}
        \left(\lasphere + 2 \musphere\right) C \left(\mat e_{rr} + \mat e_{\varphi\varphi} + \mat e_{\theta\theta}\right)
         & r < R \\
        -C \frac{4\muouter R^3}{r^3}\mat e_{rr}
        +C \frac{2\muouter R^3}{r^3}
        \left( \mat e_{\varphi\varphi} + \mat e_{\theta\theta} \right) & r > R
    \end{cases}
\end{dmath}
Now the jump condition Eq.~(\ref{eqn:jump_condition}) can be used to determine the constant $C$ which delivers the solution for the stress tensor:
\begin{dmath}
    \label{eqn:inhomo_stress}
    \mat \sigma^{\text{bulk}}
    =
    \begin{cases}
        - \dfrac{3 \lasphere + 2 \musphere}{\Lambda} \dfrac{2\intenergyab}{R}\left(\mat e_{rr} + \mat e_{\varphi\varphi} + \mat e_{\theta\theta}\right)
         & r < R \\
        \dfrac{4\muouter}{\Lambda} \dfrac{2 \intenergyab R^2}{r^3} \mat e_{rr}
        - \dfrac{2\muouter}{\Lambda} \dfrac{2 \intenergyab R^2}{r^3} \left( \mat e_{\varphi\varphi} + \mat e_{\theta\theta} \right) & r > R
    \end{cases}
\end{dmath}, 
where the shorthand $\Lambda = 4 \muouter + 3 \lasphere + 2 \musphere$ has been introduced.
For the pressure difference across the interface one obtains
\begin{dmath}
    \label{eqn:inhomo_pressure}
    \Delta p =
    \dfrac{3 \lasphere + 2 \musphere}{4 \muouter + 3 \lasphere + 2 \musphere}
    \dfrac{2 \intenergyab}{R}
\end{dmath}.

\subsection{Force balance and interface tension}
\label{app:proof}

In the following we show that the requirement of force balance at the interface Eq.~(\ref{eqn:force_balance}) is equivalent to the jump condition Eq.~(\ref{eqn:jump_condition}).
Therefore, divergences of the stress tensors $\mat \sigma_{\text{bulk}}$ and $\mat \sigma_{\text{int}}$ have to be calculated.
The divergence of torsion-free stress tensor, which depends only on $r$, can be calculated with
\begin{dmath}
    \label{eqn:div_stress_spherical}
    \nabla \cdot \mat \sigma \left(r\right) =
    \left[\frac{1}{r^2} \frac{\partial}{\partial r} \left( r^2 \sigma_{rr} \right) - \frac{\sigma_{\varphi\varphi}+\sigma_{\varphi\varphi}}{r} \right] \, \vec e_r
    + \left[\frac{\sigma_{\varphi\varphi} - \sigma_{\theta\theta}}{r}\right]\, \vec e_\varphi
\end{dmath}.
One can decompose the stress tensor of the bulk into the stress inside the inhomogeneity $\mat \sigma^{-}$ and the stress outside of it $\mat \sigma^{+}$
\begin{dmath}
    \mat \sigma^{\text{bulk}} = \mat \sigma^{-}\Theta(R-r) + \mat \sigma^{+}\Theta(r-R)
\end{dmath},
where $\Theta$ is the Heaviside function.
Since the divergence of $\mat \sigma^{-}$ and $\mat \sigma^{+}$ is zero, the divergence of $\mat \sigma^{\text{bulk}}$ delivers
\begin{dmath}
    \nabla \cdot \mat \sigma^{\text{bulk}}
    =
    \mat \sigma^{+}\delta(r-R)\, \vec e_r - \mat \sigma^{-}\delta(R-r) \, \vec e_r
\end{dmath},
where $\delta$ is the Dirac delta distribution.
The interface stress tensor $\mat \sigma^{\text{int}}$ can directly be formulated in spherical coordinates:
\begin{dmath}
    \label{eqn:stress_int}
    \mat \sigma^{\text{int}}
    =
    \intenergyab \delta \left(r-R\right) \left(\vec e_{\varphi\varphi} + \vec e_{\theta\theta}\right)
\end{dmath},
where Dirac delta distribution has been used instead of the phase-field description $I_{\alpha\beta}$ (see Eq.~(\ref{eq:idicator})).
By inserting Eq.~(\ref{eqn:stress_int}) into Eq.~(\ref{eqn:div_stress_spherical}) the force density on the surface is obtained
\begin{dmath}
    \label{eqn:force_int}
    \nabla \cdot \mat \sigma^{\text{int}} = -\frac{2\intenergyab}{r} \delta \left(r-R\right) \, \vec e_r
\end{dmath}.
One can see that the interface stress tensor results in a force which acts normal to the interface and is proportional to the curvature.
This way, the sum of force densities at the interface can be written as
\begin{dmath}
    \label{eqn:sum_force_delta}
    \mat \sigma^{+}\delta(r-R) - \mat \sigma^{-}\delta(R-r)
    = \frac{2\intenergyab}{r} \delta \left(r-R\right)
\end{dmath}.
Integrating Eq.~(\ref{eqn:sum_force_delta}) over $r$ delivers Eq.~(\ref{eqn:jump_condition}).

\section{Stress in a plate}
\label{app:plate}

Consider a finite undeformed area $A_0$ of an infinitely large plate of phase $\alpha$, the thickness $D$ and the surface tension $\sigma_{\alpha \beta}$.
Further consider the phase $\beta$ to be a dilute gas with negligible hydrostatic pressure and free of stress.
The normal vector of the plate is assumed to be along the positive $x$ direction $\vec{n}=\vec{e}_x$.
We first show that the component of the stress tensor along $\vec e_x$ is zero.
For this purpose, we write the condition of mechanical equilibrium for the $x$-component of force.
This reads
$0=f_x=\partial \sigma_{xx} /\partial x + \partial \sigma_{yx} /\partial y + \partial \sigma_{zx} /\partial z$.
Note that, for simplicity, we drop the index $\alpha$ from the stress and strain tensors ($\mat \sigma_{\alpha}=\mat \sigma$ and $\mat \varepsilon_{\alpha}=\mat \varepsilon$).
Due to homogeneity of the plate along the tangential directions $y$ and $z$, the corresponding partial derivatives vanish and one obtains $\partial \sigma_{xx} /\partial x=0$.
Thus, $\sigma_{xx}$ is constant along the $x$ direction.
Using the fact that the stress in the surrounding gas phase is negligible, one sees that this constant value must be zero: $\sigma_{xx}=0$.

Similarly, one obtains from $f_y=0$ and $f_z=0$ the important result that $\sigma_{xy}=\sigma_{yx}=0$ and $\sigma_xz=\sigma_{zx}=0$.
Due to the symmetry of the problem, it finally follows that $\sigma_{yz}=\sigma_{zy}=0$.
Thus, the stress tensor, is of a diagonal form, consisting of two equal tangential components, $\sigma_{yy}=\sigma_{zz}=\sigma_{\parallel}$, and a normal one, $\sigma_{xx}=\sigma_{\bot}$.

In order to proceed further, we simplify the elastic part of the total differential of the surface energy, Eq.~(\ref{EqdFde}), according to the present planar geometry,
\begin{dmath}\label{eqn:A1}
     \diff F =
    2 A_0 \sigma_{\alpha\beta}
    \left[ \diff \varepsilon_{I,yy} + \diff \varepsilon_{I,zz}\right] +
    A_0 D \left[\sigma_{yy} \diff \varepsilon_{yy} + \sigma_{zz} \diff \varepsilon_{zz} \right]
    \label{eqn:a}
\end{dmath}.
Furthermore, we consider only the stress in the bulk of the plate and use a sharp interface description, so that $\sigma_{yy}$ and $\sigma_{zz}$ can be considered as mere constants.
This also allows the further simplification of $\mat \varepsilon = \mat \varepsilon_I$ and $\mat \sigma = \mat \sigma_I$, so that Eq.~(\ref{eqn:a}) can be rewritten as
\begin{dmath} \label{eqn:A2}
    \diff F = 2A_0 \sigma_{\alpha\beta} \diff \varepsilon_{\parallel} +
    A_0 D \sigma_{\parallel} \diff \varepsilon_{\parallel}
\end{dmath},
where we used $\varepsilon_{\parallel}=\varepsilon_{yy}+\varepsilon_{zz}$.
Requiring that the elastic contribution to the free energy variation must be zero, $ \diff F / \diff \varepsilon_{\parallel}= 0 $, one arrives at Eq.~(\ref{eq:sigma-plate}).

\bibliographystyle{prsty}
\bibliography{interface-elasticity}

\begin{thebibliography}{10}

\bibitem{gibbs_scientific_1906}
J.~W. Gibbs, {\em The scientific papers of {J}. {Willard} {Gibbs}} (Longmans,
  Green and Company, Harlow, United Kingdom, 1906), Vol.~1.

\bibitem{shuttleworth_surface_1950}
R. Shuttleworth, Proceedings of the Physical Society. Section A {\bf 63},  444
  (1950).

\bibitem{gurtin_continuum_1975}
M.~E. Gurtin and A.~I. Murdoch, Archive for Rational Mechanics and Analysis
  {\bf 57},  291  (1975).

\bibitem{gurtin_addenda_1975}
M.~E. Gurtin and A.~I. Murdoch, Archive for Rational Mechanics and Analysis
  {\bf 59},  389  (1975).

\bibitem{cammarata_surface_1994}
R.~C. Cammarata and K. Sieradzki, Annual Review of Materials Science {\bf 24},
  215  (1994).

\bibitem{cammarata_surface_1994-1}
R.~C. Cammarata, Progress in Surface Science {\bf 46},  1  (1994).

\bibitem{gurtin_general_1998}
M.~E. Gurtin, J. Weissm{\"u}ller, and F. Larch{\'e}, Philosophical Magazine A
  {\bf 78},  1093  (1998).

\bibitem{kramer_note_2007}
D. Kramer and J. Weissm{\"u}ller, Surface Science {\bf 601},  3042  (2007).

\bibitem{bottomley_alternative_2001}
D.~J. Bottomley and T. Ogino, Physical Review B {\bf 63},  165412  (2001).

\bibitem{makkonen_misinterpretation_2012}
L. Makkonen, Scripta Materialia {\bf 66},  627  (2012).

\bibitem{yan_modified_2017}
Z. Yan and L. Jiang, Nanomaterials {\bf 7},  27  (2017).

\bibitem{sharma_effect_2003}
P. Sharma, S. Ganti, and N. Bhate, Applied Physics Letters {\bf 82},  535
  (2003).

\bibitem{sharma_size-dependent_2006}
P. Sharma and L.~T. Wheeler, Journal of Applied Mechanics {\bf 74},  447
  (2006).

\bibitem{fischer_role_2008}
F. Fischer, T. Waitz, D. Vollath, and N. Simha, Progress in Materials Science
  {\bf 53},  481  (2008).

\bibitem{slutsker_phase-field_2008}
J. Slutsker, A. Artemev, and A. Roytburd, Physical Review Letters {\bf 100},
  087602  (2008).

\bibitem{levitas_size_2011}
V.~I. Levitas and K. Samani, Nature Communications {\bf 2},  284  (2011).

\bibitem{levitas_phase_2016}
V.~I. Levitas and J.~A. Warren, Journal of the Mechanics and Physics of Solids
  {\bf 91},  94  (2016).

\bibitem{lu_dynamics_2001}
W. Lu and Z. Suo, Journal of the Mechanics and Physics of Solids {\bf 49},
  1937  (2001).

\bibitem{lu_symmetry_2002}
W. Lu and Z. Suo, Physical Review B {\bf 65},  205418  (2002).

\bibitem{lu_patterning_2004}
W. Lu and D. Kim, Nano Letters {\bf 4},  313  (2004).

\bibitem{steinbach_phase-field_2009}
I. Steinbach, Modelling and Simulation in Materials Science and Engineering
  {\bf 17},  073001  (2009).

\bibitem{steinbach_phase-field_2013}
I. Steinbach, Annual Review of Materials Research {\bf 43},  89  (2013).

\bibitem{steinbach_multi_2006}
I. Steinbach and M. Apel, Physica D: Nonlinear Phenomena {\bf 217},  153
  (2006).

\bibitem{hu_phase-field_2001}
S.~Y. Hu and L.~Q. Chen, Acta materialia {\bf 49},  1879  (2001).

\bibitem{maiya_surface_1967}
P.~S. Maiya and J.~M. Blakely, Journal of Applied Physics {\bf 38},  698
  (1967).

\bibitem{rowlinson_molecular_1982}
J.~S. Rowlinson and B. Widom, {\em Molecular {Theory} of {Capillarity}}, {\em
  The {International} {Series} of {Monographs} on {Chemistry}} (Clarendon
  Press, Oxford, Great Clarendon St, Oxford OX2 6DP, UK, 1982).

\bibitem{varnik_molecular_2000}
F. Varnik, J. Baschnagel, and K. Binder, The Journal of Chemical Physics {\bf
  113},  4444  (2000).

\bibitem{vakili_numerical_2017}
S. Vakili, I. Steinbach, and F. Varnik, Procedia Computer Science {\bf 108},
  1852  (2017).

\bibitem{schiedung_multi-phase-field_2017}
R. Schiedung, R.~D. Kamachali, I. Steinbach, and F. Varnik, Physical Review E
  {\bf 96},  012801  (2017).

\bibitem{landau_theory_1986}
L.~D. Landau and E.~M. Lifshitz, {\em Theory of {Elasticity}}, No.~7 in {\em
  Course of {Theoretical} {Physics}}, 3 ed. (Butterworth-Heinemann, Oxford,
  United Kingdom, 1986).

\end{thebibliography}

\end{document}